\documentclass[12pt,fleqn]{article}
\usepackage{epsf}
\usepackage{overcite}
\def\fnote#1#2{\begingroup\def\thefootnote{#1}\footnote{#2}\endgroup}
\makeatletter \def\@cite#1{$\m@th^{\hbox{\@ove@rcfont#1)}}$} \makeatother
\makeatletter
\def\section{\@startsection {section}{1}{\z@}{3.5ex plus 1ex minus
    .2ex}{2.3ex plus .2ex}{\sc }}
\def\subsection{\@startsection{subsection}{2}{\z@}{3.25ex plus 1ex
minus
   .2ex}{1.5ex plus .2ex}{\small \sc }}
\def\appendix{\par\clearpage
  \setcounter{section}{0}
  \setcounter{subsection}{0}
  \@addtoreset{equation}{section}
  \def\@sectname{Appendix~}
  \def\theequation{\thesection.\arabic{equation}}
  \def\thesection{\Alph{section}}}
\makeatother
\makeatletter \@addtoreset{equation}{section} \makeatother
\renewcommand{\theequation}{\thesection.\arabic{equation}}

\def\ap#1#2#3{     {\it Ann. Phys. (NY) }{\bf #1} (19#2) #3}

\def\npb#1#2#3{    {\it Nucl. Phys. }{\bf B #1} (19#2) #3}
\def\plb#1#2#3{    {\it Phys. Lett. }{\bf B #1} (19#2) #3}
\def\prd#1#2#3{    {\it Phys. Rev. }{\bf D #1} (19#2) #3}
\def\pr#1#2#3{    {\it Phys. Rev. }{\bf #1} (19#2) #3}

\def\prl#1#2#3{    {\it Phys. Rev. Lett. }{\bf #1} (19#2) #3}

\def\zpc#1#2#3{    {\it Z. Physik }{\bf C #1} (19#2) #3}

\def\nc#1#2#3{     {\it Nuovo Cim. }{\bf #1} (19#2) #3}

\def\ijmpa#1#2#3{  {\it Int. J. Mod. Phys. }{\bf A #1} (19#2) #3}
\let\vev\VEV

\def\Re{\mathop{\mbox{Re}}}

\def\etal{{\it et al.}}

\newcommand{\bea}{\begin{eqnarray}}
\newcommand{\beq}{\begin{equation}}
\newcommand{\eea}{\end{eqnarray}}
\newcommand{\eeq}{\end{equation}}

\newcommand{\spav}[1]{\parbox{1mm}{\vspace*{#1}}}

\begin{document}
\pagestyle{empty}
\begin{titlepage}
\spav{6cm}\\
\begin{center}
{\Large\bf The $\Delta I = 1/2$ Selection Rule\fnote{\dag}{To appear in the
Proceedings of the {\em Workshop
     on K Physics}, Orsay, France, May 30-June 4, 1996}} \\
\spav{1cm}\\
{\large Marco Fabbrichesi}\\
{\em  INFN, Sezione di Trieste and}\\
{\em Scuola Internazionale Superiore di Studi Avanzati}\\
{\em via Beirut 4, I-34013 Trieste, Italy.}
\vspace{6cm}\\ 
%\epsfxsize=6cm
%\centerline{\epsfbox{me.eps}}
%%
{\sc Abstract}
\end{center}
I review a recent attempt to reproduce
the isospin $I= 0$ and 2 amplitudes for the decay of a kaon
into two pions by estimating the
relevant hadronic matrix elements in the chiral quark model. The results are
parametrized in terms of  the quark and gluon condensates and of the
constituent quark mass $M$. The latter is a parameter characteristic 
of the model.
For values of  these parameters within the current
determinations,  the $\Delta I= 1/2$ selection rule is well reproduced 
by means of the cumulative effects of short-distance NLO
Wilson coefficients, penguin diagrams, non-factorizable soft-gluon 
corrections and meson-loop renormalization. 
 
\vfill
\end{titlepage}

\newpage
\setcounter{footnote}{0}
\setcounter{page}{1}

\section{Introduction}

For the decay of a neutral kaon into two
pions, the $CP$-conserving amplitude
with a final $I=0$ isospin state ($\Delta I = 1/2$)
is measured to be~\cite{PDB} 
\beq
\Re \, A_0 ( K^0 \rightarrow 2 \pi) = 3.33 \times 10^{-7} \: 
\mbox{GeV}\, , 
\eeq
and it is approximately 22 times larger than that
with the pions in the $I=2$ state ($\Delta I = 3/2$):
\beq
\Re \, A_2 ( K^0 \rightarrow 2 \pi)  = 1.50 \times 10^{-8}\: 
\mbox{GeV} \, . 
\eeq
Since a naive estimate of the relevant
hadronic matrix elements within the standard model 
leads to amplitudes that are comparable in size, this 
selection rule has been a standing
puzzle~\cite{GMP}, the solution of which has attracted a great deal 
of theoretical work 
over the past 40 years (for a review see, for instance, ref. \cite{1/2}).

While there is substantial agreement that QCD accounts for the bulk of
the rule, a quantitative (and detailed) understanding requires
showing how the different features enter. Therefore
 I will consider each of them one at the time. 
Yet, it is important to bear in mind that they are not independent effects;
 on the contrary, they are all rooted in QCD and they come about as
the scale is lowered from $m_W$ to, say, 1 GeV and the
single operator $Q_2$ develops into ten operators and gluon and
meson corrections are included.

This talk is based on ref.~\cite{ABFL}. Among the previous attempts 
in explaining the rule, I would like to
recall that in ref.~\cite{BBG} that is the closest to ours.

Let us then start from the effective lagrangian for
$\Delta S = 1$ weak transitions at the scale $\mu$:
\beq
{\cal L}_{\Delta S = 1} = 
-\frac{G_F}{\sqrt{2}} V_{ud}\,V^*_{us} \sum_i \Bigl[
z_i(\mu) + \tau y_i(\mu) \Bigr] Q_i (\mu) 
\eeq
which is defined by the ten operators
\beq
\begin{array}{rclrcl}
Q_{1} & = & \left( \overline{s}_{\alpha} u_{\beta}  \right)_{\rm V-A}
            \left( \overline{u}_{\beta}  d_{\alpha} \right)_{\rm V-A}
& \qquad
Q_{2} & = & \left( \overline{s} u \right)_{\rm V-A}
            \left( \overline{u} d \right)_{\rm V-A}
\\[1ex]
Q_{3,5} & = & \left( \overline{s} d \right)_{\rm V-A}
   \sum_{q} \left( \overline{q} q \right)_{\rm V\mp A}
& \qquad
Q_{4,6} & = & \left( \overline{s}_{\alpha} d_{\beta}  \right)_{\rm V-A}
   \sum_{q} ( \overline{q}_{\beta}  q_{\alpha} )_{\rm V\mp A}
\\[1ex]
Q_{7,9} & = & \frac{3}{2} \left( \overline{s} d \right)_{\rm V-A}
         \sum_{q} \hat{e}_q \left( \overline{q} q \right)_{\rm V\pm A}
& \qquad
Q_{8,10} & = & \frac{3}{2} \left( \overline{s}_{\alpha} 
                                                 d_{\beta} \right)_{\rm V-A}
     \sum_{q} \hat{e}_q ( \overline{q}_{\beta}  q_{\alpha})_{\rm V\pm A} \, .
\end{array} 
\eeq

\section{The Starting Point}
\begin{figure}[ht]
\epsfxsize=10cm
\centerline{\epsfbox{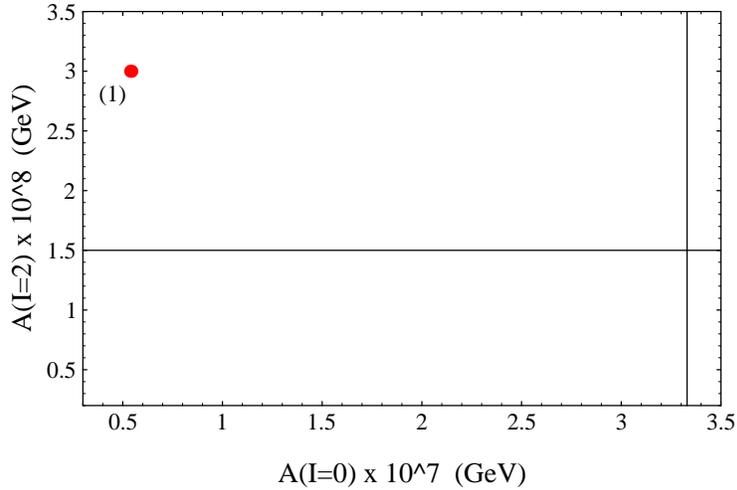}}
\caption{The starting point (1).}
\end{figure}
The starting point is computed in the absence of QCD corrections.
In this case we have only one operator, $Q_2$, and:
\beq
A_0 = \langle Q_2 \rangle _0  = \frac{2}{3} X 
 \qquad A_2 = \langle Q_2 \rangle _2  = \frac{\sqrt{2}}{3} X  
\eeq
where $X \equiv \sqrt{3} f_\pi \left( m_K^2 - m_\pi^2 \right)$.
Fig.~1 illustrates how far the amplitudes thus obtained are
from the experimental values. 

\section{NLO QCD Wilson Coefficients}
\begin{figure}[ht]
\epsfxsize=10cm
\centerline{\epsfbox{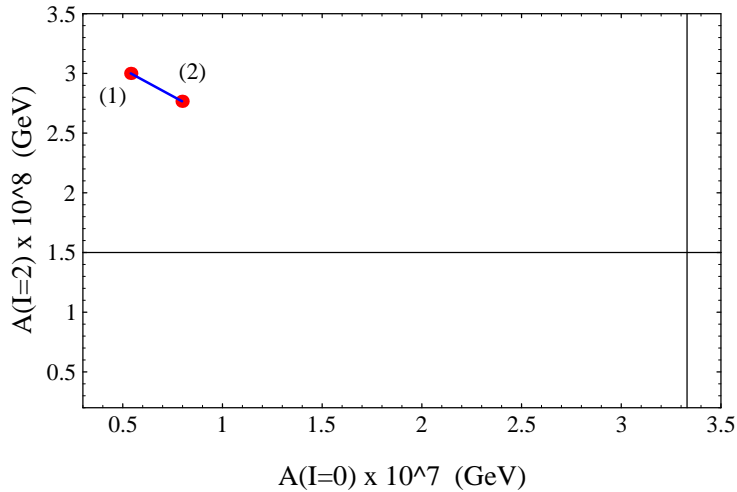}}
\caption{The effect of NLO QCD renormalization: point(2).}
\end{figure}
The first effect of having QCD in the game is the renormalization group evolution of
the Wilson coefficients~\cite{WC}. To give an idea of it, I collected
 in Table~1 the relevant next-to-leading-order (NLO) coefficients.
\begin{table}[ht]\begin{center}{\small
\begin{tabular}{|c|r r||r r|}
\hline
$\Lambda_{QCD}^{(4)}$ & \multicolumn{2}{c||}{ 250 MeV }
                      & \multicolumn{2}{c||}{ 350 MeV } \\
\hline
$\alpha_s(m_Z)_{\overline{MS}}$ 
                      & \multicolumn{2}{c||}{ 0.113 }
                      & \multicolumn{2}{c||}{ 0.119 } \\
\hline
\multicolumn{5}{c}{\mbox{(HV)}  $\quad \mu = 0.8$ GeV}\\
\hline
$z_1$&$(0.0320)$&$-0.657$&$(0.0339)$&$-0.910$ \\
\hline
$z_2$&$(0.988)$&$1.38$&$(0.987)$&$1.58$\\
\hline
$z_3$&$$&$0.0137$&$  $&$0.0301$ \\
\hline
$z_4$&$$&$-0.0292$&$  $&$-0.0540$ \\
\hline
$z_5$&$$&$0.0070$&$  $&$0.0100$ \\
\hline
$z_6$&$$&$-0.0275$&$  $&$-0.0515$ \\
\hline
$z_7/\alpha$&$$&$-0.0055$&$  $&$-0.0030$ \\
\hline
$z_8/\alpha$&$$&$0.0198$&$  $&$0.0379$ \\
\hline
$z_9/\alpha$&$$&$0.0070$&$  $&$0.0203$ \\
\hline
$z_{10}/\alpha$&$$&$-0.0181$&$  $&$-0.0330$ \\
\hline
\end{tabular}
\caption{The NLO Wilson coefficients for the ten effective operators (as 
computed in the HV scheme). The initial values at $m_W$ are between brackets.} }
\end{center}\end{table}
Fig.~2 makes clear that, while the effect goes in the right direction, it is
by far too small to account for the rule. 

\section{Chiral Quark Model}

In order to proceed we must estimate the hadronic matrix elements. 
Neither the lattice nor the $1/N_c$ approach can reproduce the rule.
Notice that the leading $1/N_c$ corrections make $A_2$ bigger and
do not help.

I will
use the chiral quark model~\cite{QM} ($\chi$QM) that is as simple a model
as it is possible without having to
renounce those features that we deem crucial in the understanding of the
rule. It is defined by the following lagrangian
\beq
 {\cal{L}}_{\chi \mbox{\scriptsize QM}} = 
- M \left( \overline{q}_R \; \Sigma
q_L +
\overline{q}_L \; \Sigma^{\dagger} q_R \right)
\eeq
that dictates the interaction between Goldstone bosons and quarks and therefore
allows us to estimate the relevant hadronic matrix elements.

\section{Penguin Operators}
\begin{figure}[ht]
\epsfxsize=10cm
\centerline{\epsfbox{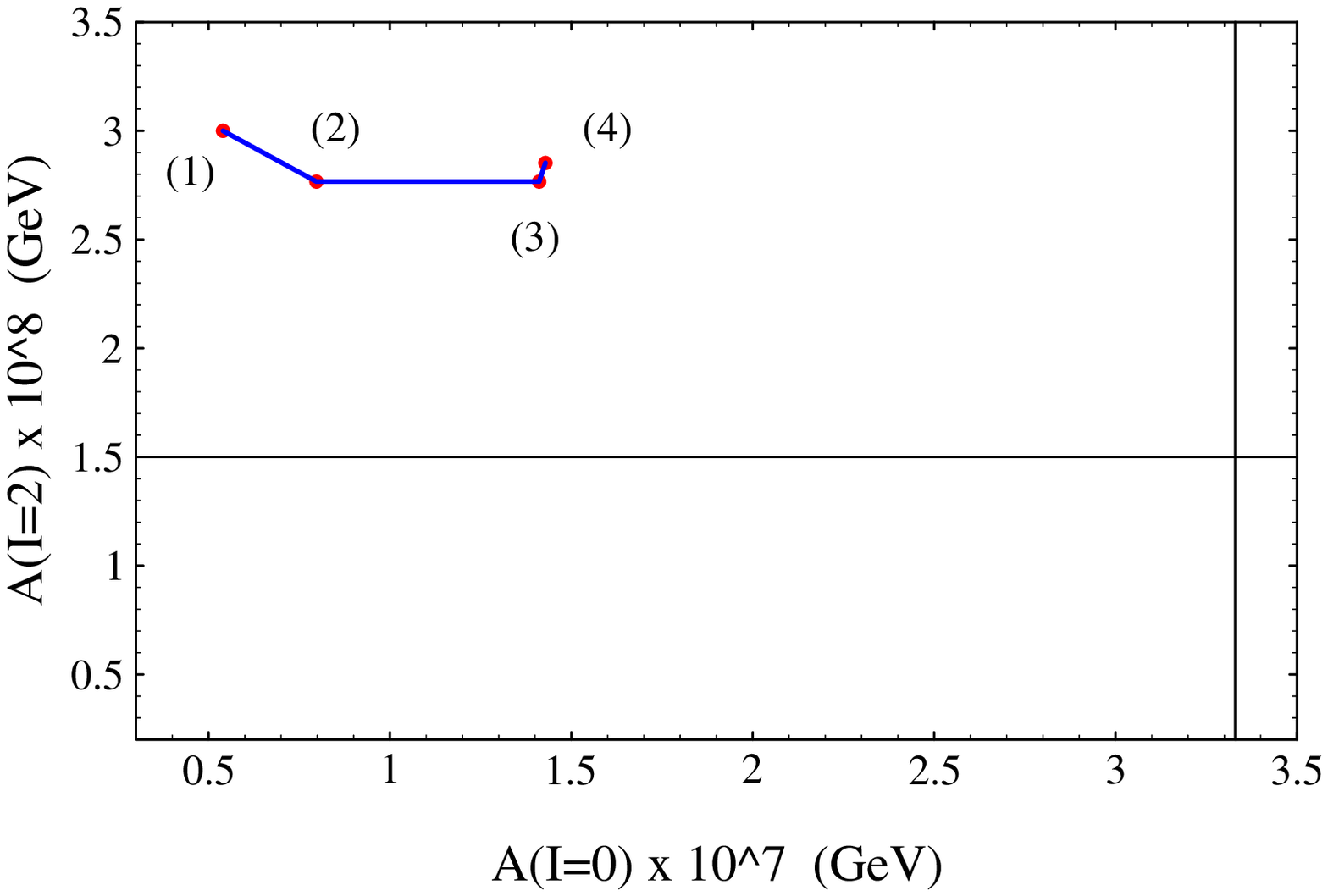}}
\caption{Penguin operators correction. Point (3) is the gluon Penguins;
point (4) the electroweak ones. $\vev{\bar q q} = (-0.250 \mbox{MeV})^3$. }
\end{figure}
A further important step in understanding the enhancement of the $A_0$ amplitude
comes from the QCD-induced penguin operators~\cite{penguin} that only affect
the $I=0$ amplitudes. In the $\chi$QM I find:
\beq
\langle Q_3 \rangle _0  =  \frac{1}{N_c} X \quad
\langle Q_4 \rangle _0  =  X  \quad
\langle Q_5 \rangle _0  =   \frac{2}{N_c}  \, 
\frac{\langle \bar{q}q \rangle}{M f_\pi^2} \, X' \quad
\langle Q_6 \rangle _0  =  2  \, \frac{\langle 
\bar{q}q \rangle}{f_\pi^2 M} \, X' \, ,
\eeq
where $X' \equiv X \left( 1 - 6\ M^2/\Lambda_\chi^2 \right)$. 
Fig.~3 shows their effect.
 
\section{Non-Factorizable Gluon Corrections}
\begin{figure}[ht]
\epsfxsize=10cm
\centerline{\epsfbox{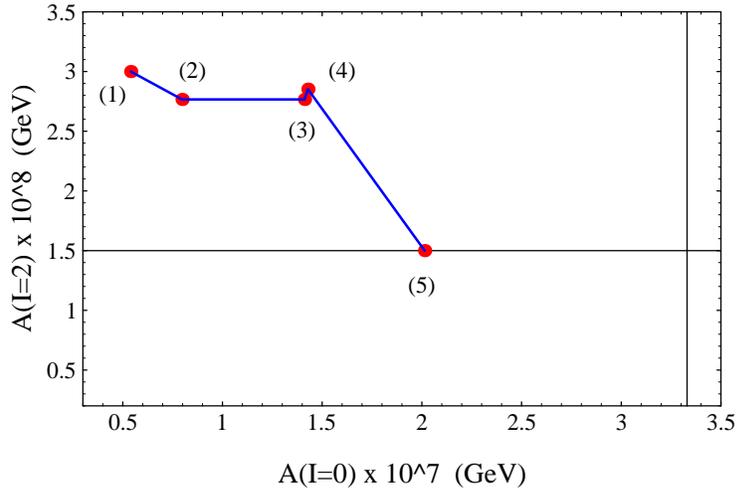}}
\caption{Nonperturbative gluon corrections: point (5).}
\end{figure}
Another, and, as we shall see essential, effect of QCD in the chiral quark
model, arises from the soft-gluon
corrections to the matrix elements~\cite{gluon}. These are parametrized by
\beq
 \delta_{\langle GG \rangle} = 
 \frac{N_c}{2} \frac{\langle 
 \alpha_s G G/\pi \rangle}{16 \pi^2 f^4} 
 \eeq
 where I take for the non-perturbative gluon condensate
 \beq
 \langle 
 \alpha_s G G/\pi \rangle = ( 350 \mbox{MeV} )^4 \, ,
 \eeq
a value that is consistent with current QCD sum rule estimates~\cite{SR}.
 The effect of this correction is 
\beq 
\begin{array}{rclrcl}
\langle Q_1 \rangle _0 & = & \frac{1}{3} X \left[ -1 + \frac{2}{N_c} \left(
1 - \delta_{\vev{GG}} \right)
\right] & \qquad
\langle Q_1 \rangle _2 & = & \frac{\sqrt{2}}{3} X \left[ 1 + \frac{1}{N_c} 
\left(
1 - \delta_{\vev{GG}} \right) 
\right] \\
\langle Q_2 \rangle _0 & = & \frac{1}{3} X \left[ 2  - \frac{1}{N_c} \left(
1 - \delta_{\vev{GG}} \right) 
\right] & \qquad
\langle Q_2 \rangle _2 & = &  \frac{\sqrt{2}}{3} X \left[ 1 + \frac{1}{N_c} 
\left( 1 - \delta_{\vev{GG}} \right) \right] \\
\langle Q_3 \rangle _0 & = & \frac{1}{N_c} X  \left(
1 - \delta_{\vev{GG}} \right) & \qquad
\langle Q_4 \rangle _0 & = & X  \\
\langle Q_5 \rangle _0 & = &  \frac{2}{N_c}  \, 
\frac{\langle \bar{q}q \rangle}{M f_\pi^2} \, X' & \qquad
\langle Q_6 \rangle _0 & = & 2  \, \frac{\langle 
\bar{q}q \rangle}{f_\pi^2 M} \, X' \, .
\end{array} 
\eeq
Fig.~4 shows how much this correction helps in going in the right direction.
The amplitude $A_2$ is brought to its experimental value.

\section{Chiral Loop Corrections}
\begin{figure}[ht]
\epsfxsize=10cm
\centerline{\epsfbox{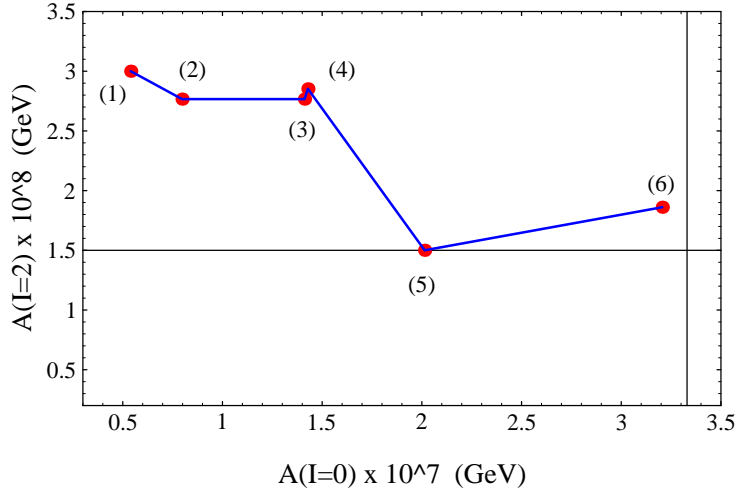}}
\caption{Effect of the chiral loops: point (6)}
\end{figure}
The amplitude $A_0$ is still too small.
Meson loops~\cite{meson} give the final enhancement. Fig.~5 shows how 
their effect is large
for the $A_0$ amplitude and small for the $A_2$, as it should be in order to 
agree with the the selection rule.

\section{The Final Point}
\begin{figure}[ht]
\epsfxsize=10cm
\centerline{\epsfbox{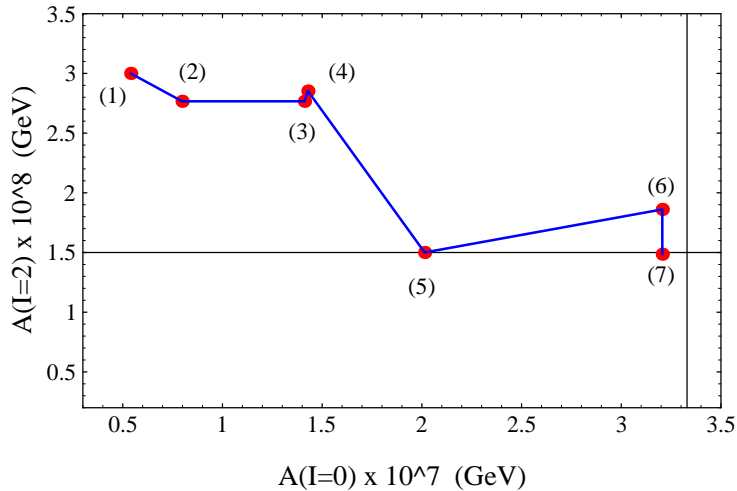}}
\caption{The final point (7).}
\end{figure}
Finally we must include  the iso-spin breaking correction to $A_2$; it is
proportional to $A_0$ and given by
\beq
A_2^{\rm iso-brk} \simeq -\frac{1}{3 \sqrt{2}} 
\frac{m_d - m_u}{m_s} A_0
\eeq
Fig.~6 includes such a correction and shows the final result. As it can be seen,
the selection rule is now well reproduced.

\section{Dependence on Input Parameters}
\begin{figure}[ht]
\epsfxsize=10cm
\centerline{\epsfbox{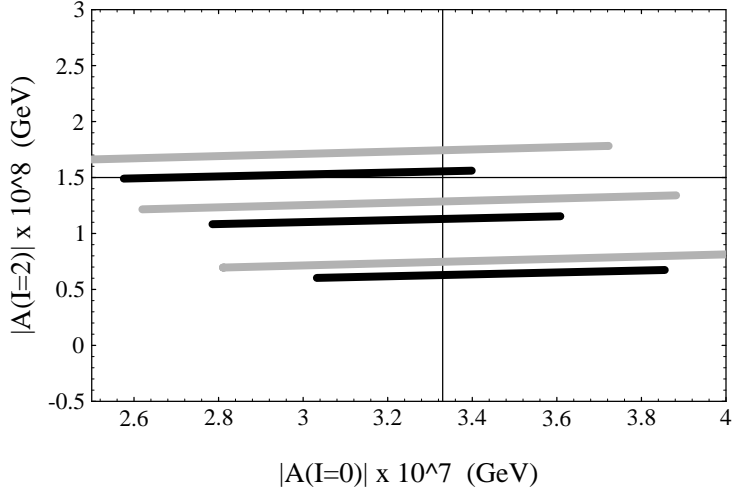}}
\caption{Input parameter dependence. Grey (black) lines are in the NDR (HV)
scheme.}
\end{figure}
While the previous figures plotted a single point 
which corresponded to a fixed value
for the input parameters, the dependence on them is rather strong as shown in
Fig.~7 where I varied the quark condensate as
\beq
- ( 200 \: \mbox{MeV} )^3 \leq 
\vev{\bar{q}q} \leq  - ( 280 \: \mbox{MeV} )^3 
\eeq
and the gluon condensate as
\beq
(346  \:\mbox{MeV} )^4 \leq \langle \frac{\alpha_s}{\pi} G G \rangle \leq
(386 \:
\mbox{MeV} )^4 
\eeq
for fixed $M = 180 \: \mbox{MeV}$ at $\mu = 0.8 \: \mbox{GeV}$. 

Because of this intrinsic uncertainty, the best strategy consists in
using the
$\Delta I = 1/2$ rule to restrict the input parameters and then use the matrix elements thus
determined to predict new physical observable like, for instance, 
$\varepsilon '/\varepsilon$~\cite{BEF}.

\section{Scale and Matching Dependence}

There is also a residual matching dependence that is described in Table~2. 
As shown by the two last lines of the table, the 
scale dependence is below the 20\% level, as opposed to that before
meson-loop renormalization which is as large as 40\%.
\begin{table}[ht]\begin{center}{\small
\begin{tabular}{|c||c|c||c|c||c|c|}
\hline
\multicolumn{7}{|c|}{$\Lambda^{(4)}_{\rm QCD}$ = 350 MeV}\\
\hline 
 & \multicolumn{2}{c||}{$\mu = 0.8$ GeV} & \multicolumn{2}{c||}{$\mu = 0.9$ GeV}
 & \multicolumn{2}{c|}{$\mu = 1$ GeV} \\
\hline
       & NDR & HV & NDR & HV & NDR & HV \\
\hline
$A_0$  & 2.97 & 2.94 & 2.66 & 2.61 & 2.45 & 2.39  \\
\hline
$A_2$  & 1.60 & 1.46 & 1.68 & 1.56 & 1.75 & 1.64  \\
\hline
$\Delta_{\gamma_5} A_0$ & \multicolumn{2}{c||}{$ 1 \%$ } & \multicolumn{2}{c||}{$ 2 \%$ } & \multicolumn{2}{c|}{$2 \%$ }\\
\hline
$\Delta_{\gamma_5} A_2$ & \multicolumn{2}{c||}{$9 \%$ } & \multicolumn{2}{c||}{$8 \%$ } &
\multicolumn{2}{c|}{$6 \%$ }\\
\hline
$\Delta_\mu A_0$ & \multicolumn{6}{c|}{$19\% -20\%$ } \\
\hline
$\Delta_\mu A_2$ & \multicolumn{6}{c|}{$9\% - 12\%$ } \\
\hline
\end{tabular}
\caption{Matching dependence.
$M \simeq 160 \: \mbox{MeV}$, $\vev{\bar q q}$ at its PCAC value and
$\vev{\alpha_s/\pi GG} = (376 \: \mbox{MeV})^4$.}}
\end{center}\end{table}

%
%------------------------- REFERENCES ------------------------------
%\clearpage
%
\renewcommand{\baselinestretch}{1}

\end{document}